\documentclass[11pt]{article}
\usepackage{subfigure}
\usepackage{color}
\usepackage{url}
\usepackage{graphicx}
\usepackage{times}
\usepackage{fullpage}

\newcommand{\qed}{\mbox{}\hspace*{\fill}\nolinebreak\mbox{$\rule{0.6em}{0.6em}$}}

\definecolor{gray}{rgb}{0.5,0.5,0.5}

\advance \topmargin by -\headheight
\advance \topmargin by -\headsep
\evensidemargin \oddsidemargin
\usepackage{hyperref}
\hypersetup{
bookmarksnumbered
}

\title{Complexity Measures for Map-Reduce, and Comparison to Parallel
Computing
  \footnote{One Disclaimer: This is an expository article, as opposed to new
    research. We are not claiming that the measures defined here have
    never been used before, nor are we claiming that our observations
    regarding the difference between PRAM and Map-Reduce are
    novel. Our motivation for writing this is two-fold: (a) We have
    often felt the need to establish a common vocabulary while
    discussing the performance of Map-Reduce algorithms with engineers
    and fellow researchers, and (b) Whenever we talk to more
    theoretical students, the question of whether Map-Reduce is just
    parallel programming in another guise always comes up.}
}
\author{Ashish Goel\thanks{Supported in part by the DARPA xdata
    program,  by grant \#FA9550-12-1-XXXX from the U.S.
Air Force Office of Scientific Research (AFOSR) and the Defense Advanced
Research Projects Agency (DARPA), and by NSF Award 0904325.}\\ Stanford University and Twitter \and Kamesh
  Munagala\\Duke University\footnote{Supported by an Alfred P. Sloan Research  Fellowship, a gift from Cisco, and
 by NSF via grants CCF-0745761, CCF-1008065, and IIS-0964560. Part of this work was done while the author was visiting Twitter, Inc.}}
\newcommand{\keyval}{\ensuremath{\langle {\mbox{\scshape{Key,
          Value}}}\rangle} }

\begin{document}
\maketitle

The programming paradigm Map-Reduce~\cite{dg:mapreduce04} and its main
open-source implementation, Hadoop~\cite{hadoop}, have had an enormous
impact on large scale data processing. Our goal in this expository
writeup is two-fold: first, we want to present some complexity
measures that allow us to talk about Map-Reduce algorithms formally,
and second, we want to point out why this model is actually different
from other models of parallel programming, most notably the PRAM
(Parallel Random Access Memory) model. We are looking for complexity
measures that are detailed enough to make fine-grained distinction
between different algorithms, but which also abstract away many of the
implementation details.

\section{An Overview of Map-Reduce}
\label{sec:mr}

Map-Reduce is commonly used to refer to both a programming model for
{\em Bulk Synchronous Parallel Processing}~\cite{v:bsp90}, as well as
a computational infrastructure for implementing this programming
model. From the infrastructure point of view, a Map-Reduce job has
three phases listed below.

While many good descriptions of Map-Reduce exist~\cite{dg:mapreduce04,
  ksv:mapreduce10}, we still would like to present a description since
one of the phases (shuffle) is typically given less attention, and
this phase is going to be crucial in our complexity measures and in
the distinction that we draw with PRAM.

\begin{description}
\item[Map:] In this phase, a User Defined Function (UDF), also called
  {\em Map}, is executed on each record in a given file. The file is
  typically striped across many computers, and many processes (called
  Mappers) work on the file in parallel. The output of each call to Map is
  a list of \keyval pairs.
\item[Shuffle:] This is a phase that is hidden from the
  programmer. All the \keyval pairs are sent to another group of
  computers, such that all \keyval pairs with the same {\scshape
    {Key}} go to the same computer, chosen uniformly at random from
  this group, and independently of all other keys. At each destination
  computer, \keyval pairs with the same {\scshape {Key}} are
  aggregated together. So if $\langle x, y_1\rangle, \langle x,
  y_2\rangle, \ldots, \langle x, y_K\rangle$ are all the key-value
  pairs produced by the Mappers with the same key $x$, at the
  destination computer for key $x$, these get aggregated into a large
  \keyval pair $\langle x, \{ y_1, y_2, \ldots, y_K\}\rangle$; observe
  that there is no ordering guarantee. The aggregated \keyval pair is
  typically called a {\em Reduce Record}, and its key is referred to
  as the {\em Reduce Key}.
\item[Reduce:] In this phase, a UDF, also called {\em Reduce}, is
  applied to each Reduce Record, often by many parallel
  processes. Each process is called a {\em Reducer}. For each
  invocation of Reduce, one or more records may get written into a
  local output file.
\end{description}

The reduce phase starts after all the Mappers have finished, and
hence, this model is an example of {\em Bulk Synchronous Processing}
(BSP).  The shuffle phase is typically implemented by writing all the
data that comes to a destination computer to disk. The task of
separating out the data into different Reduce Records on each
destination computer is also done off of disk. We are going to assume
that the total amount of work done in the shuffle phase is
proportional only to the size of the data being shuffled, both overall
as well as for any one destination computer.

\section{Complexity Measures}
\label{sec:complexity}
A good characterization of the class of problems for which the
Map-Reduce computation model can give a performance advantage over a
single machine already exists~\cite{ksv:mapreduce10}. However, our
goal here is to provide complexity measures that are sufficient to
make a fine-grained distinction between the performance of different
Map-Reduce algorithms. There are many different operations that happen
in Map-Reduce, and an exhaustive list of complexity measures such as
the one in~\cite{h:hadoop11} does not lead to easy algorithmic
analysis. We will focus on a smaller set of measures that we believe
capture essential performance bottlenecks. In particular, we keep
track of the aggregate work done by the entire system, and the work
done at the finest granularity (i.e. Mappers and Reducers)
separately. Our measures are:

\begin{description}
\item[Key Complexity:] This itself consists of three parts:
  \begin{enumerate}
  \item The maximum size of a \keyval pair input to or output by a
    Mapper/Reducer,
  \item The maximum running time for a Mapper/Reducer for a \keyval
    pair.
  \item The maximum memory used by a Mapper/Reducer to process a
    \keyval pair, and
  \end{enumerate}
\item[Sequential Complexity:] This time, we sum over all Mappers and
  Reducers as opposed to looking at the worst.
  \begin{enumerate}
  \item The size of all \keyval pairs input and output by the Mappers
    and the Reducers,
  \item The total running time for all Mappers and Reducers.
  \end{enumerate}
  Notice that we omit the total memory from our sequential complexity
  measure, since that depends on the number of Reducers operating at
  any given time and is a property of the Map-Reduce deployment as
  opposed to the algorithm.\end{description} For some problems, the
key complexity can depend on whether we assume streaming Reducers (as
in Hadoop streams~\cite{hadoopstream} which read a Reduce Record one value at
a time serially from disk or batched Reducers which take the entire
Reduce Record as input and store it in memory.

\subsection{Two Illustrative Examples and Discussion}
\label{sec:examples}
We will discuss two simple and oft-used examples, Word Count and a
single PageRank iteration, assuming the trivial Map-Reduce algorithms
in each case~\cite{dg:mapreduce04}:
\begin{description}
\item[Word Count:] Assume $N$ documents, $M$ words, total document
  size $S$, and word frequencies $f_1, f_2, \ldots, f_M$. We get the
  following complexity (assuming batched Reducers):
  \begin{itemize}
  \item Key complexity: The size, time, and memory are all
    $O(f_{MAX})$ where $f_{MAX} = \max_i f_i$.
  \item Sequential complexity: The total size and running time are
    both $O(S)$.
  \end{itemize}
  With streaming Reducers, the key complexity becomes $O(f_{MAX})$
  (size and time), and $O(1)$ (memory), whereas the sequential
  complexity remains the same.

\item[PageRank:] Given a directed graph $G=(V,E)$ with $M$ edges, $N$
  nodes, and maximum in- or out-degree $d_{MAX}$, each iteration of
  PageRank (assuming each edge is already annotated with the
  out-degree of its source node) for batched Reducers is:
  \begin{itemize}
  \item Key complexity: The size, time, and memory are all
    $O(d_{MAX})$.
  \item Sequential complexity: The total size and running time are
    both $O(M)$.
  \end{itemize}
  With streaming Reducers, the key complexity becomes $O(d_{MAX})$
  (size and time), and $O(1)$ (memory), whereas the sequential
  complexity remains the same.
\end{description}

In each of the two cases, the complexity measures are simple, and
capture natural properties of the algorithms while avoiding
implementation and deployment details.  In order to be broadly useful,
any complexity measure must capture essential aspects of the
problem. We describe several such aspects:
\begin{enumerate}
\item Our key complexity measures capture the performance of an
  idealized Map-Reduce system with infinitely many Mappers and
  Reducers, each of which can execute a single map or reduce operation on a separate machine, with no coordination overheads. In other words, a feasible Map-Reduce algorithm will  have key complexity within the typical specifications of a physical machine.
  Furthermore, a small key complexity
  guarantees that a Map-Reduce algorithm will not suffer from ``the
  curse of the last Reducer''~\cite{sv:curse11}, a phenomenon where the
  average work done by all Reducer may be small, but due to variation
  in the size of Reduce Records, the total wall clock time may be
  extremely large, or even worse, some Reducers may run out of memory.
\item The sequential complexity measures capture the total ``volume" of data generated each phase, and hence the total system
  resources consumed. In other words, this would be the amount of
  effort spent if the entire Map-Reduce installation had a single
  Mapper and a single Reducer. If the sequential complexity of a
  Map-Reduce algorithm is small (eg. if it matches the best known PRAM
  or message passing algorithm, or even better, the best known
  sequential algorithm for a problem), and the key complexity
  is small as well, then we can immediately conclude that we have an
  optimum or near-optimum Map-Reduce algorithm. Note that the total
  size input/output by all Mappers/Reducers captures the total
  filesystem I/O done by the algorithm, and is often of the order of
  the shuffle size.
\item Our complexity measures depend only on the algorithm, and
  {\em not on details of the Map-Reduce installation} such as the
  number of machines, the number of Mappers/Reducers etc., which is a
  desirable property for the analysis of algorithms. For the sake of contrast, the measures in~\cite{ksv:mapreduce10} characterize a Map-Reduce algorithm as ``good" if it uses sub-linear (in input size) number of processors each with sub-linear memory - for a graph problem,  this often forces both number of processors and memory to be $\Omega(n)$, where $n$ is the number of vertices. In contrast, our measures allow for much smaller key complexity (and hence memory requirement) by tying the performance measure to the complexity of a single key, as opposed to the complexity of work assigned to a single Mapper or Reducer. This leads the algorithm designer to make more informed trade-offs based on the hardware available.
\end{enumerate}

In our experience at Twitter, the above measures have proved to be a
valuable guide in the design of efficient Map-Reduce algorithms; while
subjective, this is arguably the ultimate test of any set of
complexity measures.

\section{PRAM vs Map-Reduce: Exploiting the Power of the Shuffle Phase}
\label{pram}

Let us consider the simple PageRank example in the PRAM model, where
the input edges reside on shared disk (to make it similar to
Map-Reduce and avoid penalizing the PRAM model for storing the $M$
edges). If we have $K$ machines, then the total I/O and the total
running time are both $O(M)$, which are matched by
Map-Reduce. However, the total memory needed by all the PRAM machines
is $O(N)$ where $N$ is the number of nodes, and the memory needed by
each PRAM machine is $O(N/K)$. In Map-Reduce with $K$ Reducers, by
contrast, the total memory needed by all Reducers (assuming streaming
Reducers) is $O(K)$ and the memory needed by each Reducer is
$O(1)$, assuming the Reducer processes values for one key after it is completely done processing values for another key. 
Similar differences exist in the even simpler Word Count
example.

This seems surprising, and on first glance, might appear to be a flaw
in our modeling. However, we believe this gets exactly to one of the
reasons why Map-Reduce is so successful as a computational paradigm
(beyond the obvious ease-of-use reasons). In Map-Reduce, the shuffle
phase aggregates all the \keyval pairs into Reduce Records, and the
cost for this step is not being charged to the algorithm by our
complexity measures. This accurately captures the practical design of
Map-Reduce platforms: the shuffle phase first writes everything onto
disk at each destination machine, and then aggregates the received
\keyval pairs into Reduce Records. The writing on disk is something
that needs to happen anyway because of the BSP model, and dominates
the cost of the aggregation phase. Hence, by using the disk as
temporary memory, Map-Reduce isolates the cost of aggregation from
system performance\footnote{Of course, if we were to take disk usage
  into account, a Map-Reduce algorithm would use more memory than the
  PRAM model.}.

The difference in memory usage can be substantial (eg. for small $K$),
and hence, designing efficient Map-Reduce algorithms is an interesting
research question in its own right, {\em distinct from the design of
  efficient PRAM algorithms}. Many of the algorithms that we are
currently working on (eg.~\cite{disco}) exploit the fact that we get
the aggregation step for free as part of a shuffle.

\medskip
Another point of difference with a PRAM is that we are separating out the number of phases and the reduce key complexity. The latter quantity could involve sequential computation, but as long as the magnitude of this computation is bounded and reasonable to execute on one physical processor, we separate it from the number of phases. In other words, this model allows a trade-off of the form: $O(\sqrt{n})$ key complexity, and $O(\log n)$ phases to process input of size $n$. The PRAM model only captures the extreme case where all computation is parallel, where it would appear that the parallel running time for the above example is $O(\sqrt{n} \log n)$. It is therefore conceivable that the Map-Reduce model efficiently solves problems that do not have efficient parallel algorithms in the traditional sense.

\medskip
While not germane to this article, we would like to point out another
important reason behind the success of Map-Reduce. In modern systems,
the network is much faster than disk, but the network is a shared
resource. By having many Mappers and Reducers, the same shared network
bandwidth drives many disks during the shuffle phase. Just like
writing to disk in the shuffle phase hides the cost of aggregation,
using a shared network hides the cost of disk accesses. It would be
very interesting to see how large-scale adoption of faster solid state
disks changes this equation.

\section{The Aggregation Exception}
It is also important to point out that there are several aspects of
typical Map-Reduce systems that we do not model, most notably the {\em
  Combine} operation, which is like running a Reducer locally at each
Mapper. The combine operation is the most beneficial for aggregation
operations, where we need to apply a simple operator such as sum, max,
or average to all Map Records. To capture the benefit of Combination,
we need to introduce the number of Mappers, $K$, into our complexity
measures. This gives the following complexity for sum/max/average and
many similar aggregation functions, where $N$ is the number of Map
Records (assuming batched Reducers):
\begin{itemize}
\item Key complexity: The size, time, and memory are all
  $O(K)$.
\item Sequential complexity: The total size and running time are
  both $O(N)$.
\end{itemize}
With streaming Reducers, the key complexity becomes $O(K)$ (size and
time), and $O(1)$ (memory). In practical installations, $O(K)$ is
typically negligible compared to the coordination overhead in a
Map-Reduce phase. Hence, we recommend just treating the key complexity
as $O(1)$ for these operations.

For researchers who disagree with our recommendation (or where the
distinction is important in the problem), using $K$ explicitly in the
complexity measures is a reasonable alternative.

\pdfbookmark[1]{\refname}{My\refname}
\bibliographystyle{plain}
\bibliography{global}

\begin{thebibliography}{1}

\bibitem{hadoop}
http://hadoop.apache.org.

\bibitem{hadoopstream}
http://wiki.apache.org/hadoop/hadoopstreaming.

\bibitem{dg:mapreduce04}
Jeffrey Dean and Sanjay Ghemawat.
\newblock Mapreduce: Simplified data processing on large clusters.
\newblock In {\em OSDI}, pages 137--150. USENIX Association, 2004.

\bibitem{h:hadoop11}
Herodotos Herodotou.
\newblock Hadoop performance models.
\newblock {\em CoRR}, abs/1106.0940, 2011.

\bibitem{ksv:mapreduce10}
H.~Karloff, S.~Suri, and S.~Vassilvitskii.
\newblock A model of computation for mapreduce.
\newblock {\em Proceedings of the Twenty-First Annual Symposium on Discrete
  Algorithms (SODA '10)}, 2010.

\bibitem{sv:curse11}
S.~Suri and S.~Vassilvitskii.
\newblock Counting triangles and the curse of the last reducer.
\newblock {\em Proceedings of the 20th International Conference on World Wide
  Web (WWW '11)}, pages 607--614, 2011.

\bibitem{v:bsp90}
Leslie~G. Valiant.
\newblock A bridging model for parallel computation.
\newblock {\em Commun. ACM}, 33(8):103--111, 1990.

\bibitem{disco}
Reza~Bosagh Zadeh and Ashish Goel.
\newblock Dimension independent similarity computation.
\newblock {\em CoRR}, abs/1206.2082, 2012.

\end{thebibliography}
\end{document}